\documentclass[twocolumn]{aa}

\usepackage{graphicx}
\usepackage{txfonts}
\usepackage{color}

\begin{document}

\title{Non-equilibrium ionization states in galaxy clusters.}

\author{D. A. Prokhorov \inst{1,2,3}}

\offprints{phdmitry@gmail.com}

\institute{UPMC Universit\'e Paris 06, UMR~7095, Institut
d'Astrophysique de Paris, F-75014, Paris, France \and CNRS,
UMR~7095, Institut d'Astrophysique de Paris, F-75014, Paris, France
\and
Korea Astronomy and Space Science Institute, 61-1 Hwaam-dong,
Yuseong-gu, Daejeon 205-348, Korea}

\date{Accepted . Received ; Draft printed: \today}

\authorrunning{Prokhorov D. A.}

\titlerunning{Non-equilibrium ionization}

\abstract {X-ray imaging observatories have revealed hydrodynamic
structures with linear scales $\sim 10$ kpc in clusters of galaxies,
such as shock waves in the 1E0657-56 and A520 galaxy clusters and
the hot plasma bubble in the MKW 3s cluster. The future X-ray
observatory IXO will resolve for the first time the metal
distribution  in galaxy clusters at the these scales.} {Heating of
plasmas by shocks and AGN activities can result in non-equilibrium
ionization states of metal ions. We study the effect of the
non-equilibrium ionization at linear scales $\lesssim 50$ kpc in
galaxy clusters. } {A condition for non-equilibrium ionization is
derived by comparing the ionization time-scale with the age of
hydrodynamic structures. Modeling of non-equilibrium ionization when
the plasma temperature suddenly change is performed. An analysis of
relaxation processes of the FeXXV and FeXXVI ions by means of
eigenvectors of the transition matrix is given.} {We conclude that
the non-equilibrium ionization of iron can occur in galaxy clusters
if the baryonic overdensity $\delta$ is smaller than $11.0/\tau$,
where $\tau\ll1$ is the ratio of the hydrodynamic structure age to
the Hubble time. Our modeling indicates that the emissivity in the
helium-like emission lines of iron increases as a result of
deviation from the ionization equilibrium. A slow process of
helium-like ionic fraction relaxation was analyzed. A new way to
determine a shock velocity is proposed.}{}


\maketitle

\section{Introduction}

Clusters of galaxies are gravitationally bound structures of mass
$\sim 10^{14}-10^{15}$ M$_{\bigodot}$ and size $\sim 1-3$ Mpc (for
a review, see \cite{Kaastra 2008} 2008). Their mass budget
consists of dark matter $(\simeq 80\%)$, hot diffuse intracluster
plasma $(\lesssim 20\%)$ and a small fraction of other components
such as stars and dust. The mean baryonic overdensity in galaxy
clusters equal to
$\delta=n_{\mathrm{H}}/\bar{n}_{\mathrm{H}}\simeq 200$, where
$n_{\mathrm{H}}$ and $\bar{n}_{\mathrm{H}}$ are the mean hydrogen
densities in galaxy clusters and in the Universe, respectively.

Many chemical elements reside in galaxy clusters. The plasma
temperatures kT$\simeq 3-10$ keV in galaxy clusters are close to
the values of the K-shell ionization potentials of heavy elements
($I_{Z}=Z^2 Ry$, where Z is the atomic number and $Ry$ the Rydberg
constant). Emission lines from heavy elements were detected by
X-ray telescopes from galaxy clusters. The current instruments
(XMM-Newton, Chandra and Suzaku) have largely enhanced our
knowledge on the chemical abundances of many elements. Metal
abundances around 0.3 in Solar Units of \cite{Anders 1989} (1989)
were derived under the assumptions of collisional ionization
equilibrium (for a review, \cite{Werner 2008} 2008).

Non-equilibrium processes such as non-equilibrium ionization and
relaxation of the ion and electron temperatures are usually taken
into account only in the outskirts of galaxy clusters and in the
warm hot intergalactic medium (WHIM) where baryonic overdensity
$\delta$ is less than 200 (e.g. \cite{Yoshikawa 2006} 2006;
\cite{Prokhorov 2008} 2008). However, we show that non-equilibrium
ionization can also be produced as the result of merging processes
and AGN activity in galaxy clusters where the baryonic overdensity
$\delta \gtrsim 1000$.

Evidences for merging processes of galaxy clusters and AGN
activity, such as shocks and hot plasma bubbles, were revealed by
means of Chandra high-resolution observations. For example, strong
shocks in the 1E0657-56 and A520 galaxy clusters propagating with
a velocity of 4700 km/s and 2300 km/s, respectively, were derived
by \cite{Markevitch 2002} (2002) and \cite{Markevitch 2005}
(2005). The corresponding Mach numbers of the shocks are $3.0$ and
$2.1$. Hot plasmas inside bubbles arising from AGN activity were
detected in galaxy clusters (e.g. \cite{Mazzotta 2002} 2002).
Heating of plasmas produced by shocks or AGNs can result in a
non-equilibrium ionization state.

In this paper we study the effect of non-equilibrium ionization
near merger shock fronts and in hot plasma bubbles. We give a
theoretical analysis of collisional
non-equilibrium ionization in Sect.~2. We show numerically the
importance of this effect in galaxy clusters in Sect.~3. We
analyze helium-like and hydrogen-like non-equilibrium ionization
states by means of eigenvectors of the transition matrix in
Sect.~4. A new approach to determine the value of the shock velocity
is considered in Sect.~5 and our results are discussed in Sect.~6.

\section{A condition for non-equilibrium ionization.}

Non-equilibrium ionization is often assumed in supernova remnants
(e.g. \cite{Gronenschild 1982} 1982, \cite{Masai 1994} 1994) and
may be important in the WHIM (\cite{Yoshikawa 2006} 2006).
The non-equilibrium ionization state in the linked region between
the Abell 399 and Abell 401 clusters was also studied by \cite{Akahori
2008} (2008). We are going to show that non-equilibrium ionization can occur
not only in the outskirts of galaxy clusters but also in galaxy
clusters, in which merging processes and AGN activity play a
role. We derive here a condition on the baryonic overdensity for
deviation from collisional ionization equilibrium, by comparing
the ionization time-scale with the age of hydrodynamic structures.

The number of collisions between electrons and an ion resulting in
electron impact ionization per unit time is
$\nu=\sigma_{\mathrm{ion}} v_{\mathrm{thr}} \tilde{n}$, where
$\sigma_{\mathrm{ion}}$ is the characteristic value of the
ionization cross-section, $v_{\mathrm{thr}}$ corresponds to the
threshold velocity (energy) of the ionization process, and
$\tilde{n}$ is the number density of electrons which have
sufficient energy for electron impact ionization. The ionization
time-scale is given by $t_{\mathrm{ion}}=1/\nu$, and therefore
\begin{equation}
t_{\mathrm{ion}}=\frac{1}{\sigma_{\mathrm{ion}} v_{\mathrm{thr}}
\tilde{n}}.\label{tion}
\end{equation}

As it was noted by \cite{Yoshikawa 2006} (2006) and \cite{Akahori
2008} (2008), helium-like and hydrogen-like ions are interesting
for the analysis of non-equilibrium ionization. Therefore, electron
impact ionization of a helium-like ion will be considered in this
section as a physically important case (a consideration of
hydrogen-like ions is analogous).

For helium-like ions the characteristic value of the ionization
cross-section is approximately (see \cite{Bazylev 1981} 1981)

\begin{equation}
\sigma^{\mathrm{He}}_{\mathrm{ion}}\approx\frac{2 \pi
a^2_{\mathrm{0}}}{Z^4}\label{sigma}
\end{equation}
where $a_{\mathrm{0}}=\hbar^2/m_{\mathrm{e}} e^2$ is the Bohr
radius, Z is the atomic number.

Since the ionization potential of a He-like ion is approximately
$I_{\mathrm{Z}}\approx Z^2 m_{\mathrm{e}} e^4
\hbar^{-2}/2$\footnote{The exact value of the ionized potential of
a He-like ions is obtained by changing $Z$ with
$Z_{\mathrm{eff}}=Z-5/16$. We consider the case $Z\gg1$ and,
therefore, the approximative value is sufficient.}, the electron
threshold velocity $v_{\mathrm{thr}}=\sqrt{2
I_{\mathrm{Z}}/m_{\mathrm{e}}}$ can be approximated as
\begin{equation}
v_{\mathrm{thr}}\approx Z \frac{e^2}{\hbar}. \label{v}
\end{equation}

The number density of electrons with energies higher than the
ionization potential of a He-like ion is

\begin{equation}
\tilde{n}= n_{0} \int^{\infty}_{\mathrm{p_{\mathrm{thr}}}} x^2
f_{\mathrm{M}}(x) dx \label{tilden}
\end{equation}
where $n_{0}$ is the plasma number density,
$p_{\mathrm{thr}}=\sqrt{2 I_{\mathrm{Z}}/kT}$ is the dimensionless
threshold momentum and
$f_{\mathrm{M}}(x)=\sqrt{2/\pi}\times\exp\left(-x^2/2\right)$ is
the Maxwellian distribution.

If the dimensionless threshold momentum $p_{\mathrm{thr}}\gtrsim
1$ then a simplified form of Eq. (\ref{tilden}) is given by

\begin{equation}
\tilde{n}\approx \frac{2}{\sqrt{\pi}} n_{0}
\exp\left(-\frac{I_{\mathrm{Z}}}{kT}\right)
\sqrt{\frac{I_{\mathrm{Z}}}{kT}}\label{tildenew}
\end{equation}

Using Eqs. (\ref{sigma}), (\ref{v}) and (\ref{tildenew}) we
rewrite Eq. (\ref{tion}) as

\begin{equation}
t_{\mathrm{ion}}\approx \frac{1}{4\sqrt{\pi}} \frac{Z^2
m^{3/2}_{\mathrm{e}} \sqrt{kT}}{n_{0} \hbar^2}
\exp\left(\frac{I_{\mathrm{Z}}}{kT}\right)
\end{equation}

It is most convenient to write the plasma number density in terms
of the baryonic overdensity $n_{\mathrm{0}}=\delta
\Omega_{\mathrm{b}} \rho_{\mathrm{crit}}/m_{\mathrm{p}}$, where
the critical density is
$\rho_{\mathrm{crit}}=3H^2_{\mathrm{0}}/(8\pi G)$, and to denote
the ratio of the thermal energy $kT$ and the ionization potential
by $\lambda=kT/I_{\mathrm{Z}}$. Thus,

\begin{equation}
t_{\mathrm{ion}}\approx \frac{2\sqrt{\pi}}{3\sqrt{2}} \frac{Z^3
m^{2}_{\mathrm{e}} m_{\mathrm{p}} e^2 G}{\hbar^3
H^{2}_{\mathrm{0}}\Omega_{\mathrm{b}}}
\frac{\Phi(\lambda)}{\delta}
\end{equation}
where $\Phi(\lambda)=\sqrt{\lambda}\exp\left(1/\lambda\right)$.

Ionization states will be non-equilibrium if the ionization
time-scale $t_{\mathrm{ion}}$ is longer than the hydrodynamic
structure age $t = \tau H^{-1}_{\mathrm{0}}$, i.e.
$t_{\mathrm{ion}} > \tau H^{-1}_{0}$. This condition is equivalent
to the inequality

\begin{equation}
\delta \lesssim \frac{2\sqrt{\pi} Z^3 m^{2}_{\mathrm{e}}
m_{\mathrm{p}} e^2 G}{3\sqrt{2}\hbar^3
H_{\mathrm{0}}\Omega_{\mathrm{b}}} \frac{\Phi(\lambda)}{\tau}.
\label{delta}
\end{equation}

In an important case of the iron ions (Z=26), the numerical value
of the first dimensionless term on the right-hand side of Eq.
(\ref{delta}) is

\begin{equation}
\frac{2\sqrt{\pi} Z^3 m^{2}_{\mathrm{e}} m_{\mathrm{p}} e^2
G}{3\sqrt{2} \hbar^3 H_{\mathrm{0}}\Omega_{\mathrm{b}}}\approx 2.4
\end{equation}

and, therefore,

\begin{equation}
\delta \lesssim 2.4 \frac{\Phi(\lambda)}{\tau}.
\end{equation}

In rich galaxy clusters with plasma temperature of
$kT_{\mathrm{pl}}\approx 5$ keV, the value of the function
$\Phi(kT_{\mathrm{pl}}/I_{\mathrm{Z}})$ is $\approx4.6$.
Therefore, in this case we find $\delta \lesssim 11.0/\tau$ and
conclude that, if the hydrodynamic structure age is of order
$10^{7}-10^{8}$ years (i.e. $\tau$ lies in the range
$6.6\times10^{-4}<\tau<6.6\times10^{-3}$) then non-equilibrium
ionization occurs in galaxy clusters where the baryonic
overdensity $\delta \approx 1000$ (see Eq. \ref{delta}).

\cite{Mazzotta 2002} (2002) have estimated the age of a hot
plasma bubble of diameter $\sim 50$ kpc to be
$t\approx 2.7\times10^7$ yr, which is much shorter than the age of the MKW 3s
cluster. In the 1E0657-56 and A520 clusters the downstream
velocities of the shocked gas flowing away from the shock are 1600
km s$^{-1}$ and 1000 km s$^{-1}$ (\cite{Markevitch 2002} 2002,
\cite{Markevitch 2005} 2005), therefore the shocked gas covers
a distance $50$ kpc in times $3.1\times10^7$ and $4.7\times10^7$
yrs respectively. Thus, in light of the above conclusion
non-equilibrium ionization can occur at linear scales $\lesssim$
50 kpc in galaxy clusters in which merging processes and
AGN activity present.

\section{Modeling of non-equilibrium ionization.}

Non-equilibrium ionization occurs when the physical conditions of
the plasma, such as the temperature, suddenly change. Shocks, for
example, can lead to an almost instantaneous rise in temperature
and to a deviation from ionization equilibrium. However, it takes
some time for the plasma to respond to an instantaneous
temperature change, as the ionization balance is recovered by
collisions.

In this section we consider the following situation: the plasma
temperature instantaneously increases from $kT_{1}=3.4$ keV to
$kT_{2}=10.0$ keV. Such a temperature change may correspond to a
temperature jump at a shock with a Mach number $M=2.6$ or to plasma
heating by AGN activity. We assume that the age of the hot plasma
region is $3\times10^7$ yr and the baryonic overdensity is
$\delta=4000$ which corresponds to the plasma number density in the
post-shock region in the A520 cluster (see Fig. 2b of Markevitch et
al. 2005). Following \cite{Markevitch 2006} (2006) we assume that
the electron and ion temperatures are equal.

\begin{figure}[h]
\centering
\includegraphics[angle=0, width=8cm]{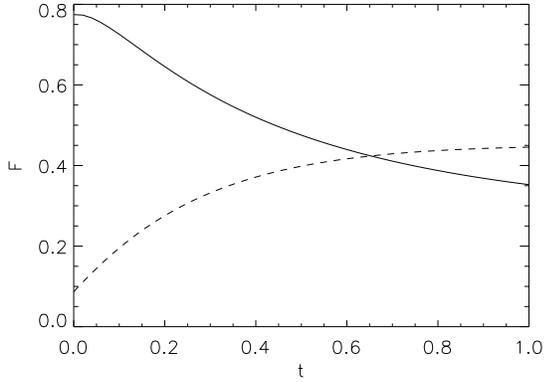}
\caption{Dependence of the He-like (solid line) and H-like (dashed
line) ionic fractions of iron on the dimensionless time
$t/(3\times10^7$ yr$)$.} \label{fractions}
\end{figure}

At the temperature $kT_{1}=3.4$ keV the ionic fractions of Li-like,
He-like and H-like ions of Fe relative to the total Fe abundance are
$\approx 12\%$, $\approx 77\%$, and $\approx 9\%$ respectively.
Therefore we consider below four iron-ion-states Fe(+23), Fe(+24),
Fe(+25) and Fe(+26). In the case the collisional ionization rate
equation for each element is written as

\begin{eqnarray}
\frac{d}{n_{0}dt}\vec{n}=\left(\begin{array}{cccc}-I_{+23} & R_{+24} & 0 & 0\\
I_{+23} & -I_{+24}-R_{+24} & R_{+25} & 0\\
0 & I_{+24} & -I_{+25}-R_{+25} & R_{+26}\\
0 & 0 & I_{+25} & -R_{+26}\end{array}\right)\vec{n} \label{system}
\end{eqnarray}
where $\vec{n}$ is the vector with four components ($n_{+23}$,
$n_{+24}$, $n_{+25}$, $n_{+26}$), normalized such that
$\sum^{26}_{\mathrm{i}=23} n_{+\mathrm{i}}=1$, which correspond to
the four iron-ion-states mentioned above, $I_{+\mathrm{z}}$ and
$R_{+\mathrm{z}}$ represent the rate coefficients for ionization and
recombination from an ion of charge z to charges z+1 and z-1,
respectively. All the coefficients necessary to calculate the direct
ionization cross sections are taken from \cite{Arnaud 1985} (1985),
the radiative recombination rates are taken from \cite{Verner 1996}
(1996), and the dielectronic recombination rates are taken from
\cite{Mazzotta 1998} (1998). To solve the system of equations
(\ref{system}) we use the fourth order Runge-Kutta method.

The time-dependence of the He-like and H-like ionic fractions of iron
is shown in Fig. \ref{fractions}.

At the temperature $kT_{2}=10.0$ keV the equilibrium ionic
fractions of He-like and H-like ions of iron are $\approx 27\%$
and $\approx 45\%$ respectively. Therefore, the He-like ionic
fraction which equals 35\% at time $3\times10^7$ yr does not reach
its equilibrium value and non-equilibrium ionization occurs.
However, the H-like ionic fraction almost reaches its equilibrium
value at time $3\times10^7$ yr.

We now show that the effect of non-equilibrium ionization on the
helium-like emission lines of iron can be significant and that
non-equilibrium ionization leads to the increase of volume
emissivity in the helium-like spectral lines.

The helium-like volume emissivity for a chemical element of atomic
number Z is given by

\begin{equation}
\epsilon_{\mathrm{Z}}=n_{\mathrm{e}} n_{\mathrm{H}}
A_{\mathrm{Z}}\times\left(n_{+\mathrm{(Z-2)}}
Q_{\mathrm{+(Z-2)}}+n_{+(\mathrm{Z-1})}
\alpha_{\mathrm{+(Z-1)}}\right)
\end{equation}
where $n_{\mathrm{e}}$ is the electron number  density,
$n_{\mathrm{H}}$ is the hydrogen number density, $A_{Z}$ is the
abundance of the considered chemical element, $n_{+(\mathrm{Z-2})}$
and $n_{+(\mathrm{Z-1})}$ are the ionic fractions of helium-like and
hydrogen-like ions respectively, $Q_{\mathrm{+(Z-2)}}$ is the impact
excitation rate coefficient and $\alpha_{\mathrm{+(Z-1)}}$ is the
rate coefficient for the contribution from radiative recombination
to the spectral lines. Excitation rate coefficients are taken from
\cite{Prokhorov 2009} (2009). Let us note the reduced volume
emissivity in the iron helium-like emission lines as

\begin{equation}
U=\frac{\epsilon_{\mathrm{Z=26}}}{n_{\mathrm{e}} n_{\mathrm{H}}
A_{\mathrm{Z}} \Gamma}
\label{epsilon}
\end{equation}
where $\Gamma=Z^{-4} \pi a^2_{0}
\sqrt{I_{\mathrm{Z}}/m_{\mathrm{e}}}$ corresponds to the
characteristic rate coefficient value (see also \cite{Prokhorov
20092} 2009).

\begin{figure}[h]
\centering
\includegraphics[angle=0, width=8cm]{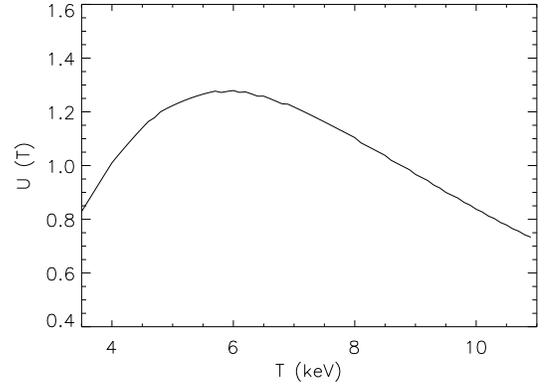}
\caption{Dependence of the equilibrium reduced iron  volume
emissivity in the helium-like lines on the plasma temperature.}
\label{U}
\end{figure}

In Fig. {\ref{U}} the reduced emissivity $U$ when the ionic fractions
are in ionization equilibrium is shown in the range of
temperatures between 3.5 keV and 11 keV.

In the situation considered above we find that at temperatures
$kT_{1}=3.4$ keV and $kT_{2}=10.0$ keV the equilibrium values of
the reduced volume emissivities are approximately equal. However,
in the presence of non-equilibrium processes approximate equality
of these volume emissivities does not hold. Since the fraction of
electrons with energy higher than the impact excitation threshold
$E_{\mathrm{ex}}\approx6.7$ keV is $27\%$ at the temperature
$kT_{1}=3.4$ keV and is much less than $72\%$ that is at
temperature $kT_{2}=10.0$ keV, more effective impact excitation
should be at temperature $kT_{2}=10.0$ keV. Furthermore the
non-equilibrium ionic fraction of helium-like iron in the region
of temperature $kT_{2}=10.0$ keV is higher than the equilibrium
ionic fraction (see Fig.\ref{fractions}) and, therefore,
non-equilibrium ionization leads to the increase of volume
emissivity in the helium-like spectral lines.

\begin{figure}[h]
\centering
\includegraphics[angle=0, width=8cm]{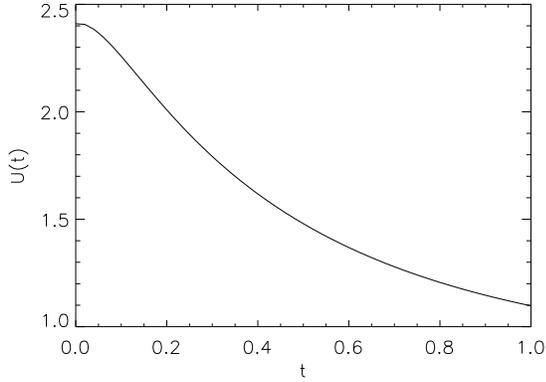}
\caption{Evolution of the reduced iron volume emissivity in the
iron helium-like lines in the region with temperature 10.0 keV.
The dimensionless time is given by t/($3\times10^{7}$ yr).}
\label{time evolution}
\end{figure}

Using the dependence of the ionic fractions of iron on the
dimensionless time $t/(3\times10^7$ yr$)$ (see Eq. \ref{system})
we study the time evolution of the reduced volume emissivity in
the iron helium-like emission lines. This time evolution is shown
in Fig. \ref{time evolution}.

Since the maximal value $U_{\mathrm{max}}$ of the equilibrium
reduced emissivity $U(T)$ is at the temperature $kT\approx 6$ keV
and equal to $\approx 1.25$ (see Fig. \ref{U}), we conclude that
the non-equilibrium value of $U(t)$ (see Fig. \ref{time
evolution}) is higher than the maximal equilibrium value
$U_{\mathrm{max}}$ up to the time $t=2\times10^7$ yr (up to the
dimensionless time equal to 0.65). The value $U_{\mathrm{eq}}$ of
the equilibrium reduced emissivity at the temperature $kT= 10.0$
keV equal to $\approx0.8$ is less than the non-equilibrium value
of $U(t)$ during the time interval $t=3\times10^7$ yr. Therefore,
the iron abundance $A_{\mathrm{Z, eq}}$ derived from the
assumption that ionization states are in equilibrium will be
higher than the correct iron abundance value $A_{\mathrm{Z}}$ (see
Eq. \ref{epsilon}), which is given by
\begin{equation}
A_{\mathrm{Z}}= A_{\mathrm{Z, eq}} \frac{U_{\mathrm{eq}}}{U(t)}.
\end{equation}

\section{An analysis of He-like and H-like non-equilibrium ionization
states by means of eigenvectors}

In the previous section we showed that the ionic fraction of
He-like iron ions can remain in non-equilibrium while the H-like iron
ionic fraction almost achieves equilibrium. This
somewhat paradoxical behavior can be more easily understood by
means of eigenvectors of the transition matrix M, which is (see
Eq. \ref{system})

\begin{equation}
M=n_{0} t_{\mathrm{age}}\left(\begin{array}{cccc}-I_{+23} & R_{+24} & 0 & 0\\
I_{+23} & -I_{+24}-R_{+24} & R_{+25} & 0\\
0 & I_{+24} & -I_{+25}-R_{+25} & R_{+26}\\
0 & 0 & I_{+25} & -R_{+26}\end{array}\right), \label{trmat}
\end{equation}
where $t_{\mathrm{age}}=3\times10^{7}$ yr is the age of the
hydrodynamical structure (see Sect.~3).

Here we calculate the values of the eigenvalues of the transition
matrix and the corresponding eigenvectors, and show how the ionic
fraction of He-like iron ions can remain non-equilibrium longer
than that of H-like iron ions.

The eigenvalues $\lambda$ of the transition matrix M are derived
from the equation
\begin{equation}
\mathrm{Det}(M-\lambda\times E)=0,
\end{equation}
where $E$ is the unit matrix.

One of the eigenvalues of the transition matrix M is of the
form Eq. (\ref{trmat}) equal to zero ($\lambda_{0}=0$).
Consequently, the ionization equilibrium is achieved in the end.

The solution of the system of differential equations (Eq.
\ref{system}) can be written as

\begin{equation}
\vec{n}=\sum^{3}_{\mathrm{i=0}}c_{\mathrm{i}}\vec{V}_{\mathrm{i}}
\exp\left(\frac{\lambda_{\mathrm{i}}t}{t_{\mathrm{age}}}\right),
\label{n}
\end{equation}
where $c_{\mathrm{i}}$ are constants, $\vec{n}$ is the vector
$(n_{+23}, n_{+24}, n_{+25}, n_{+26})$ and $\vec{V}_{\mathrm{i}}$
are the eigenvectors of the transition matrix M.

At the temperature $kT=10.0$ keV we derive three eigenvalues which
equal to $\lambda_{1}\approx -17.80$, $\lambda_{2}\approx-3.40$
and $\lambda_{3}\approx-1.46$. The eigenvectors which correspond
to the derived eigenvalues are respectively
\begin{equation}
\vec{V_1}=\left(\begin{array}{c} -0.66\\ 0.74\\ -0.08\\
0.00\end{array}\right), \ \
\vec{V_2}=\left(\begin{array}{c}-0.03\\ -0.48\\ 0.82\\
-0.31\end{array}\right), \ \
\vec{V_3}=\left(\begin{array}{c}-0.03\\ -0.61\\ -0.14\\
0.78\end{array}\right).
\end{equation}

The eigenvector which corresponds to the eigenvalue $\lambda_{0}$
determines equilibrium ionic fractions at temperature $kT=10.0$
keV.

Since $\lambda_{3}$ is the smallest absolute value of the
eigenvalues (excluding $\lambda_{0}$ which does not correspond to
any relaxation process) the process which corresponds to the
eigenvector $\vec{V_{3}}$ is the slowest (see Eq. \ref{n}). This
slow process corresponds to the increase in the FeXXVII ionic
fraction due to decreases in the FeXXV and FeXXVI ionic fractions.
However, the absolute value of the second component of
$\vec{V}_{3}$, which corresponds to the decrease in the FeXXV
ionic fraction and equals 0.61, is higher than the absolute
value of the third component of $\vec{V}_{3}$, which corresponds
to the decrease in the FeXXVI ionic fraction and equals 0.14.
Therefore, the variation in the helium-like FeXXV ionic fraction
which is proportional to the value of the second component of
$\vec{V}_{3}$ is more substantial during this relaxation process
than the variation in the hydrogen-like FeXXVI ionic fraction.

\begin{figure}[h]
\centering
\includegraphics[angle=0, width=8cm]{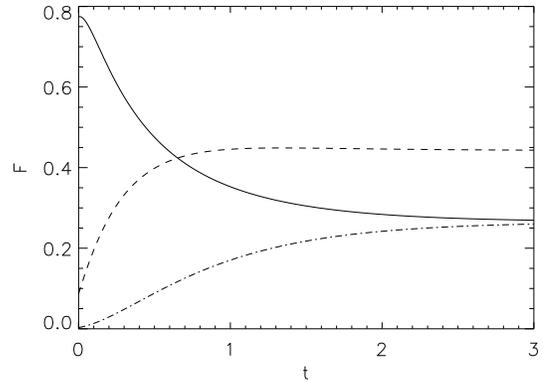}
\caption{Dependence of the He-like (solid line), H-like (dashed
line) and fully ionized (dot-dashed line) ionic fractions of iron
on the dimensionless time $t/(3\times10^7$ yr$)$.}
\label{3fractions}
\end{figure}

The quantitative difference between the variations in the He-like
and H-like ionic fractions is illustrated in Fig.
{\ref{3fractions}}.

\section{Mach number determination}

Clusters of galaxies form via gravitational infall and mergers of
smaller mass concentrations. In the course of a merger, a
significant portion of the kinetic energy of the colliding
subclusters, that carried by the gas, is dissipated by shocks.

The shocks in the A520 and 1E0657-06 clusters have Mach
numbers $M=2-3$, derived from the Rankine-Hugoniot jump
conditions (for a review, see \cite{Markevitch 2007} 2007), relating
the density and temperature jumps at the shock and the Mach
number, $M=v/c_{1}$, where $c_{1}$ is the velocity of sound in the
pre-shocked gas and $v$ is the velocity of the pre-shock gas in
the reference frame of the shock. Thus, if the pre-shock and
post-shock temperatures ($T_{1}$ and $T_{2}$, respectively) are
determined from observations, the Mach number of the shock can be
derived from the equation:

\begin{equation}
\frac{T_{2}}{T_{1}}=\frac{(2\gamma
M^2_{\mathrm{sh}}-(\gamma-1))\times((\gamma-1)
M^2_{\mathrm{sh}}+2)}{(\gamma+1)^2 M^2_{\mathrm{sh}}}
\end{equation}
where $\gamma$ is the adiabatic index. It is usually assumed that
the pre-shock velocity in the reference frame of the shock is
equal to the shock velocity in the reference frame of the galaxy
cluster and that the adiabatic index is $\gamma=5/3$ (see \cite{Markevitch
2007} 2007).

The comparison of the X-ray image and  gravitational lensing mass
map of the 1E0657-06 merging cluster (\cite{Clowe 2006} 2006)
shows that the mass peak of the subcluster is offset from the
baryonic mass peak. \cite{Clowe 2006} (2006) interpret this as the
first direct evidence for the existence of dark matter.

Such merging clusters offer the unique opportunity to study gas
physics through direct comparison of the observed shock properties
with the predictions of gas + dark matter modeling (e.g.
\cite{Prokhorov 2007} 2007; \cite{Springel 2007} 2007). In this
section we provide a new way to derive shock parameters
based on measurements of the flux ratio of the FeXXV and FeXXVI
iron lines.

The fluxes of the FeXXV and FeXXVI lines have the same dependence
on the metal abundance, as well as on the emission measure, their
ratio is independent of these parameters. This iron line ratio can
therefore be used to determine the temperature of the intracluster
gas (e.g. \cite{Nevalainen 2003} 2003) and the presence of
supra-thermal electrons (e.g. \cite{Prokhorov 2009} 2009).

Taking into account both electron-impact-excitation and radiative
recombination the iron line flux ratio is given by
\begin{equation}
R=\frac{n_{+24} Q^{1-2}_{\mathrm{FeXXV}} +
n_{+25}\alpha^{1-2}_{\mathrm{RR}\, \mathrm{FeXXV}}}
{n_{\mathrm{+25}}Q^{1-2}_{\mathrm{FeXXVI}} +
n_{+26}\alpha^{1-2}_{\mathrm{RR},\ \mathrm{FeXXVI}} },
\end{equation}
where the rate coefficients are
$Q^{1-2}_{\mathrm{FeXXV}}=\sum\limits_{a}\sum\limits_{b (<a)}
S^{1s^2-a}_{\mathrm{FeXXV}} B_{ab}$,
 $Q^{1-2}_{\mathrm{FeXXVI}}=\sum\limits_{a}\sum\limits_{b
(<a)} S^{1s-a}_{\mathrm{FeXXVI}} B_{ab}$, and
$S^{1s-a}_{\mathrm{FeXXVI}}$ are the impact-excitation rates.  The
excited states $b$ correspond to the upper levels of the He-like
triplet and the H-like doublet, and the radiative, branching
ratios are given by,
\begin{equation}
B_{ab}=\frac{A_{ab}}{\sum\limits_{c (<a)} A_{ac}} \; .
\end{equation}
and $\alpha^{1-2}_{\mathrm{RR},\ \mathrm{FeXXV}}$ and
$\alpha^{1-2}_{\mathrm{RR},\ \mathrm{FeXXVI}}$ are the rate
coefficients for the contribution from radiative recombination to
the spectral lines FeXXV (He-like triplet) and FeXXVI (H-like
doublet), respectively, and $A_{ac}$ are the transition
probabilities.

\begin{figure}[]
\centering
\includegraphics[angle=0, width=8cm]{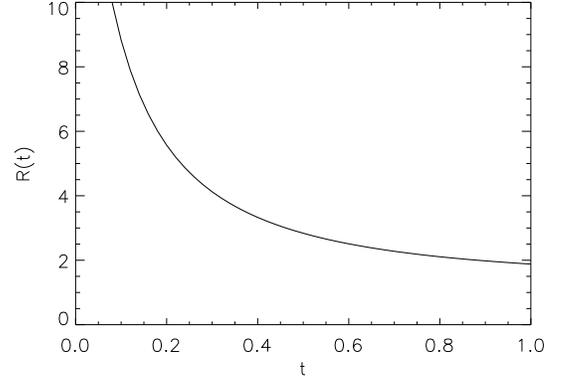}
\caption{Evolution of the iron line flux ratio in the region with
temperature 10.0 keV. The dimensionless time is given by
t/($3\times10^{7}$ yr).} \label{ironlineratio}
\end{figure}

Below we study the situation which was considered in Sects. 3 and
4. The variation of the iron line flux ratio in the region with
temperature 10.0 keV as a function of dimensionless time is shown
in Fig. \ref{ironlineratio}.

Let the downstream velocity of the shocked gas which flows away
from the shock be $V_{\mathrm{d}}$. Then the distance covered by
shocked gas passes in a time $t$ is $L = V_{d} t$. Therefore, if
the iron line flux ratio $R$ is known from observations at
distance $L$ from the shock front then using the function $R(t)$
we can derive the value of the downstream velocity $V_{d}=L/t(R)$,
where $t(R)$ is the inverse function for $R(t)$.

The Mach number of the shock and the downstream velocity are
related by (e.g. \cite{Landau 1959} 1959)
\begin{equation}
M_{\mathrm{sh}}=\sqrt{\frac{2+(\gamma-1) M^2_{\mathrm{d}}}{2\gamma
M^2_{\mathrm{d}}-(\gamma-1)}} \label{Msh}
\end{equation}
where $M_{\mathrm{d}}=V_{\mathrm{d}}/c_{2}$ and $c_{2}$ is the
velocity of sound in the post-shocked gas.

On the observational side, it will be important to derive the
flux ratio of the FeXXV and FeXXVI iron lines from the region
between the shock front and the considered distance $L$ which
shocked gas covers in a time $t = L/V_{\mathrm{d}}$. Since
non-equilibrium ionization can occur at linear scale $\lesssim 50$
kpc (see Sect. 2), we choose $L=25$ kpc. The flux ratio of the
iron lines FeXXV and FeXXVI from this region is then

\begin{equation}
R=\frac{\int^{L/V_{\mathrm{d}}}_{0} \left(n_{+24}
Q^{1-2}_{\mathrm{FeXXV}} + n_{+25}\alpha^{1-2}_{\mathrm{RR}\,
\mathrm{FeXXV}}\right)\times dt} {\int^{L/V_{\mathrm{d}}}_{0}
\left(n_{\mathrm{+25}}Q^{1-2}_{\mathrm{FeXXVI}} +
n_{+26}\alpha^{1-2}_{\mathrm{RR},\ \mathrm{FeXXVI}}\right)\times
dt },\label{R}
\end{equation}

\begin{figure}[]
\centering
\includegraphics[angle=0, width=8cm]{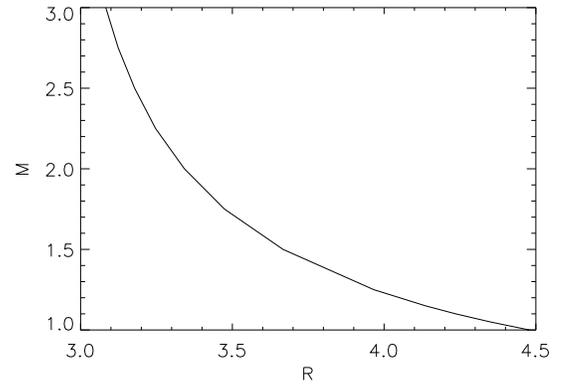}
\caption{Dependence of the Mach number of the shock M on the iron
line flux ratio R.} \label{mr}
\end{figure}

Using Eqs. (\ref{Msh}) and (\ref{R}), we find the Mach number $M$
of the shock as a function of the iron line flux ratio $R$. The
dependence M(R) is plotted in Fig. \ref{mr}.

Thus, we conclude that the Mach number of the shock can be derived
from the iron line flux ratio. Methods based on the
Rankine-Hugoniot jump conditions and on measuring the iron line
flux ratio are independent for deriving the Mach number of the
shock.

\section{Conclusions}

The currently operating X-ray imaging observatories provide us
with a detailed view of the intracluster medium in galaxy
clusters. Chandra's 1$^{\prime\prime}$ angular resolution, the
best among the current X-ray observatories, corresponds to linear
scales $< 1$ kpc at $z<0.05$ and $\approx 4$ kpc at $z=0.3$ (the
redshift of the 1E0657–56 cluster). This enables us to study
hydrodynamic phenomena in galaxy clusters, such as shock waves and
hot plasma bubbles.

Metal observations are always limited by the number of X-ray
photons. For diffuse low surface brightness objects, like galaxy
clusters, the effective area is a major issue, therefore for a
reasonable observation time, Chandra metal abundance maps will have
lower spatial resolution than XMM-Newton (see \cite{Werner 2008}
2008). IXO \footnote{http://ixo.gsfc.nasa.gov/} is planned to be a
follow-up mission of XMM-Newton and will have a sensitivity much
higher than XMM-Newton. The expected effective area of the IXO
mirror and focal plane instruments showing the large improvement at
all energy (including the 6-7 keV band) in comparison with those of
current X-ray observatories is plotted in a figure
\footnote{http://ixo.gsfc.nasa.gov/images/science/effective-area.jpg}.
The larger effective area in the 6-7 keV band the higher accuracy of
the iron line flux measurements is achieved. With IXO we will be
able to resolve for the first time the metal distribution in the ICM
on the scales of single galaxies in nearby clusters (simulations of
metallicity maps {{which will be provided by the next generation
X-ray telescope}} are given by \cite{Kapferer 2006} 2006).
Therefore, the sensitivity of IXO will provide metal observations
near shock fronts and in hot plasma bubbles.

We have considered in this paper the non-equilibrium ionization at
linear scales $\lesssim 50$ kpc in galaxy clusters. The necessary
condition on the baryonic overdensity (see Eq.  \ref{delta}) for
the existence of non-equilibrium ionization in regions of galaxy
clusters where $\delta>200$ holds for ions of iron. This is
because the iron atomic number Z=26 is high enough for the
threshold value of the overdensity, which is proportional to
$Z^{3}$ (see Eq. \ref{delta}), to become higher than the mean
cluster baryonic overdensity.

The dependence of the He-like and H-like ionic fractions of iron
on time is given in Sect.~3. We found that the He-like ionic
fraction of iron does not achieve its equilibrium value during the
age of the hydrodynamical structures and non-equilibrium
ionization takes place.

We calculated the reduced emissivity in the He-like iron spectral
lines and concluded that the iron abundance derived from the
assumption that ionization states are in equilibrium predicted to
be higher than the correct iron abundance value (see
Fig.~\ref{time evolution}).

We found that the slowest relaxation process corresponds to the
increase in the FeXXVII ionic fraction due to decreases in the
FeXXV and FeXXVI ionic fractions. However, the decrease in the
FeXXV ionic fraction is much higher than the decrease in the
FeXXVI ionic fraction during this relaxation process.

A new way to derive the Mach number of a shock based on measurements
of the flux ratio of the FeXXV and FeXXVI iron lines is proposed in
Sect.~5. The advantage of this method with respect to the method
based on the Rankine-Hugoniot jump conditions is that the first is
more accurate. Fortunately the iron line flux ratio is constrained
without the effect of hydrogen column density ($N_{\mathrm{H}}$)
uncertainties. In practice, the X-ray data can be fitted in a narrow
band containing the FeXXV and FeXXVI lines, where the absorption is
negligible (see \cite{Nevalainen 2009} 2009). The drawback is that
the number of photons is small in this narrow energy band, but the
next generation X-ray telescope IXO with larger effective area
overcomes this drawback and will be able to measure the flux ratio
of the iron $K_{\mathrm{\alpha}}$ lines and, therefore, the Mach
number of a shock with high precision. Using the narrow energy band
instead of the full X-ray spectrum minimizes the dependence on
calibration accuracy (see \cite{Nevalainen 2003} 2003), therefore
the FeXXV to FeXXVI lines are insensitive to the details of the
effective area function in contrast to the continuum spectrum. Note
that the method based on the Rankine-Hugoniot jump conditions uses
the densities and temperatures derived from the continuum spectrum.

Another advantage of the proposed method is that it permits us to
determine independently the Mach number of a shock by using
measurements of the iron line flux ratio at different distances from
a shock (see Sect. 5) since it takes into account an evolution of
ionization states.

The effect of the apparent iron overabundance under the assumption
of ionization equilibrium and the slow process of helium-like ionic
fraction relaxation should be analyzed in galaxy clusters by means
of future X-ray observatories and may have implications in different
astrophysical plasmas (e.g. in supernova remnants). New
high-spectral-resolution instruments with higher sensitivity, such
as IXO, are needed to measure the flux ratio of the iron K$\alpha$
lines with the purpose of independent by determining the shock
parameters.

\begin{acknowledgements}
I am grateful to Joseph Silk, Florence Durret, Igor Chilingarian and
Anthony Moraghan for valuable suggestions and discussions and thank
the referee for very useful comments.
\end{acknowledgements}

\end{document}